\documentclass{llncs}

\usepackage{amsmath}
\usepackage{subfigure}
\usepackage{oz}
\usepackage{refcal}
\usepackage{url}
\usepackage{rcp}

\usepackage{xcolor}
\DeclareFontShape{OT1}{cmr}{bx}{sc}{<-> cmbcsc10}{}

\usepackage{datetime}
\usepackage{rcoms}

\usepackage{xspace} 

\newcommand{\kinote}[1]{{\color{red!10!green!20!blue!95!}#1}}
\newcommand{\FR}{\textsc{Flush{\raisebox{0.4ex}{$\scriptscriptstyle+$}}Reload}\xspace}
\newcommand{\ER}{\textsc{Evict{\raisebox{0.4ex}{$\scriptscriptstyle+$}}Reload}\xspace}

\usepackage{opsem}
\usepackage{wmm}
\usepackage{caches}

\renewcommand{\asgn}{\mathbin{:\!=}}

\renewcommand{\rowsl}{{\tt IMP-ro}\xspace}
\renewcommand{\crowsl}{{\tt IMP-ro-spec}\xspace}

\renewcommand{\globals}[2]{(\keywordfont{global}~#1 @ #2)}

\renewcommand{\Update}[3]{#1{_{[{#2} \smallasgn {#3}]}}}

\renewcommand{\plasgnsmall}{\mathop{{\mbox{\scriptsize \T{+}}=}}}

\newcommand{\labeleqn}[1]{\label{eqn:#1}}
\newcommand{\labelsect}[1]{\label{sect:#1}}
\newcommand{\labellaw}[1]{\label{law:#1}}

\newcommand{\refsect}[1]{Sect.~\ref{sect:#1}}
\newcommand{\refeqn}[1]{(\ref{eqn:#1})}
\newcommand {\reffig}[1] {Fig.~\ref{fig:#1}}

\newcommand {\refappendix}[1] {Appendix~\ref{appendix:#1}}

\newcommand{\privatedata}{\textsc{d}}
\newcommand{\privatevar}{\textsc{k}}
\newcommand{\privateaddr}{\getAddr{\privatevar}}
\newcommand{\privateaddrA}{\chi}

\renewcommand{\ruledefNamed}[4]{
	\begin{minipage}[t]{#1}
	\begin{OpRule}[#2]
	\label{rule:#3}
	~ \\
	$
	#4
	$
	\\
	$~ $
	\\
	$~ $
	\vskip 0.5mm
	\end{OpRule}
	\end{minipage}
}

\title{
An abstract semantics of speculative execution for reasoning about security vulnerabilities
}
\institute{
Defence Science and Technology Group, Australia, and
\\
School of Information Technology and Electrical Engineering \\ University of Queensland
}
\author{
Robert J. Colvin \and Kirsten Winter
}

\begin{document}

\maketitle

\begin{abstract}

Reasoning about correctness and security of software is increasingly difficult due to the complexity of modern microarchitectural features such as out-of-order execution.  A class of
security vulnerabilities termed Spectre that exploits side effects of speculative, out-of-order execution was announced in 2018 and has since drawn much attention.  In this paper we
formalise speculative execution and its side effects with the intention of
allowing speculation to be reasoned about abstractly at the program level, limiting the exposure to processor-specific or low-level semantics. 
To this end we encode and expose speculative execution explicitly in the programming language, rather than solely in the operational semantics; as a result the effects of
speculative execution are captured by redefining the meaning of a conditional statement, and introducing
novel language constructs that model transient execution of an alternative branch.
We add an abstract cache to the global state of the system, and derive some general refinement rules that expose cache side effects due to speculative loads.  
Underlying this extension is 
a semantic model that is based on instruction-level parallelism.
The rules are encoded in a simulation tool, which we use to analyse an abstract specification of a Spectre attack and vulnerable
code fragments.



\OMIT{
\kinote{Caches have been modelled previously in sequentially consistent behaviour.
We now combine a cache model with a formal model of WMM semantics (i.e., extend our WMM operational semantics with a notion of caches).
Based on  this we can simulate Spectre-like attacks using Maude to confirm the behaviour of the extended operational semantics.
This operational semantics will be used to extend the existing type system for information flow security on WMM
to allow the user to check whether speculative execution behaviour causes information leaks through the cache that
currently go unnoticed. 
On the other hand, this semantics provides a base for simulation tools that allow the user to
experiment with new attack strategies exploiting WMM in general and speculative execution in particular.
}
}

\end{abstract}

\section{Introduction}
\labelsect{intro}

Modern multicore architectures exhibit several features to speed up execution: commands may appear to occur out of order, allowing
computation to proceed past some bottleneck (e.g., loading a value from memory), several levels of faster intermediate memory (caches) to speed up repeated accesses,
and in particular, \emph{speculative execution}, where a branch is optimistically executed, even though local computation may not yet have determined if it is the
correct branch.  Such features are difficult to reason about, though there has been significant work in understanding weak memory models 
\cite{x86-TSO,alg11,UnderstandingPOWER,HerdingCats,ModellingARMv8,PromisingSemantics} and also detailed formal microarchitectural models (e.g., \cite{OpSemCacheCoherent}).

Recently several significant security vulnerabilities have been found related to out-of-order execution, \eg, Meltdown \cite{lip2018},
Foreshadow \cite{bulck2018}, and Spoiler \cite{islam2019spoiler}.  In this paper we focus on the recently published Spectre class of attacks \cite{spectre2019,koc2018}.
Spectre differs in that the attack may target the victim's code to retrieve private information, while other attacks exploit processor features only.
While complex to exploit, Spectre is a vulnerability present in almost all modern architectures. 
It allows malicious code to access the memory of a victim process,
potentially reading private data, without sharing the virtual memory space.
The attack works by detecting footprints in the cache left by speculative execution; for instance, a branch that includes a bounds check on an index $i$ into an array
$A$ may speculatively load the element at $A[i]$, before it knows for certain that $i$ is within the bounds of $A$.  Though the speculative computations leading
up to the point where the mis-speculation is detected are discarded, depending on the subsequent access patterns there may still be 
an effect on the cache, which is not discarded, and which can be used to infer the value in memory at out-of-bounds address $A[i]$.

\OMIT{
Formalisations of weak memory model semantics so far (\cite{x86-TSO,alg11,UnderstandingPOWER,HerdingCats,ModellingARMv8,PromisingSemantics,col18})
ignore the effects of speculations and the cache behaviour 
since their results are discarded (apart from those stored in the cache) when speculated along the wrong branch. 
As soon as we consider the cache as being a visible entity this omission is unsatisfactory. 
}

In earlier work we have proposed a semantic framework to support reasoning about weak memory models \cite{FM18}
which is implemented in a simulation/model checking tool based on Maude \cite{Maude}.
In this paper we extend this framework with a model of cache behaviour and speculative execution.  Although Spectre may
occur in memory models that provide sequential consistency, a weak memory model framework is a natural fit for speculative execution as
speculated instructions may begin out of order, i.e., before the
relevant branch is reached. This enables not only a close inspection of Spectre-like attacks but also 
the analysis of other related potential vulnerabilities that may arise in modern hardware architectures.
Our intention for the semantics is to allow analysis of vulnerability to Spectre-like attacks to be integrated within a more general, software-level reasoning framework;
we do not aim to precisely model the
implementation of speculative execution or caches for a particular architecture.

\OMIT{
The key part of the semantics is to lift a load to become a load plus its side effect on the cache.
}

\OMIT{
The Spectre class of attacks is broader than this, and hence a general framework for formally analysing victim code and the effects of speculative execution is
needed for formal analysis.  For instance, 
Spectre can be thought of as an information flow problem \cite{mur16,mur18}, with potentially sensitive data moving from a secure
domain to an insecure domain.
To reason about the vulnerability of code to Spectre-like attacks requires a formal semantics 
that takes cache behaviour as well as speculative execution into account. 
alongside with the effects of weak memory model behaviours.
formal techniques that enable reasoning about cache side effects alongside traditional safety \kinote{and security} concerns would be advantageous.
}

Speculative execution presents several challenges. Firstly, it requires a model of the cache which, 
for our concerns, needs to be modelled at a level that presents enough details to realistically capture 
the effects of speculative execution, but is abstract enough to not over-complicate reasoning.
Secondly, speculative execution should allow side effects to take effect before the relevant branch is reached.
Thirdly, speculation can be nested, and the target of future branches may depend on speculatively executed computations, necessitating the creation of \emph{transient}
state that can be easily discarded.
Finally, we want to be able to model
and explore possible mitigations, e.g., memory barriers to halt speculation, such as Intel's \T{LFENCE} instruction \cite[Sect. 11.4.4.3]{Intel64ArchManual}.

\OMIT{From a security perspective the ideal outcome of modelling work is the identification of further vulnerabilities; theoretically our model
provides the potential to explore victim code snippets.
More generally the semantics, considering reordering in weak memory models, should be able to to be
incorporated into existing techniques for reasoning about programs, including information flow analysis as mentioned above.}

The paper is structured as follows: in \refsect{background} we summarise a wide-spectrum language and its semantics for reasoning about weak memory models.  In \refsect{crowsl} 
we extend this with new constructs for reasoning about speculative execution, and give its semantics.
We formalise some attacker and victim patterns, in particular those of Spectre, in \refsect{spectre}.  Related work is discussed in \refsect{relwork}.

\section{Background: \wmml}
\labelsect{background}

A wide-spectrum language for reasoning about weak memory models, \wmml, is introduced in \cite{FM18,rowsl-arXiv}.
It is essentially an imperative language with assignments, conditionals and loops, with the
difference that instead of sequential composition ($c_1 \scomp c_2$ for $c_i$ a command) it has prefixing, $\aca \cbef \cmdc$, where $\aca$ is an instruction
(as in process algebras such as CSP \cite{CSP} and CCS \cite{CCS}).
The semantics of prefixing is defined so that either $\aca$ may be executed, or some instruction $\acb$ from within $\cmdc$ may be executed, 
provided that $\acb$ can be
\emph{reordered} before $\aca$ according to the rules of the memory model.  To instantiate \wmml for a particular memory model a ``reordering relation'' $\ro$ on instructions is 
defined, stating when instructions can occur out of order; in addition, different models may also have different instruction types, for instance, memory barriers for
enforcing order.

We recap \wmml below, before extending it to include speculative execution in later sections.  We ground our work in a weak memory framework because speculation can occur before 
preceding instructions are executed, even when speculative execution is implemented on architectures which enforce sequential consistency. 
In addition, it appears that increasingly security vulnerabilities will be found due to instruction
reordering on modern architectures, e.g., \cite{islam2019spoiler}.  However, the particular
reordering relation is not important for the analysis in this paper, and to avoid distraction we mostly assume sequential consistency.  

The elements of \wmml are actions (instructions) $\alpha$, commands (programs) $c$, processes (local state and a command) $p$, and the top level system $s$,
encompassing a shared state and all processes.
We assume a set of variables $Var$, divided into locals (registers) and globals.
By convention we use $r, r_1, r_2$, etc., to name local variables, and unless otherwise stated, $x,y,z$ for global variables.
A state $\sigma$ is a mapping from $Var$ to values, with the notation $\Update{\sigma}{x}{v}$ representing an update of $\sigma$ to map $x$ to $v$.
Below $x$ is a variable (shared or local) and $e$ an expression.
\begin{equation*}
\begin{aligned}
	\aca &\ttdef 
		x \asgn e
		\csep
		\guarde
	\\
	\cmdc &\ttdef
		\Skip 
		\csep
		\aca \cbef \cmdc
		\csep
		\aca \bef \cmdc
		\csep
		\cmdc_1 \choice \cmdc_2
		\csep
		\WHbc
	\\
	\prp &\ttdef
		\locals{\sigma}{\cmdc}
	\\
	\syss &\ttdef
		\globals{\sigma}{\prp_1 \pl \prp_2 \pl \ldots}
\end{aligned}
\end{equation*}

An action may be an update $x \asgn e$ or a guard $\guarde$.
For weak memory models the set of actions may also include fences (memory barriers); we introduce an abstract barrier in later sections.
Commands include the terminated command $\Skip$, prefixing, choice, and iteration.  
We also include the abstract command type for ``true prefixing'', $\aca \bef \cmdc$, where reordering is forbidden, \ie, $\bef$ is prefixing
in the usual CSP \cite{CSP} and CCS \cite{CCS} sense.
For brevity, for a command $\aca \cbef \acb \cbef \Skip$ we omit the trailing $\Skip$ and just write $\aca \cbef \acb$.
A process encapsulates a command within a local state $\sigma$ (total on local variables), representing registers.
A system is structured as the parallel composition of processes sharing a global state, each with their own values for local variables.

\begin{figure}[t]

\ruledefNamed{80mm}{Prefix}
{reorder-rule-2}{
	\prefixac \tra{\aca} \cmdc
	~~~(a)
	\qquad
	\Rule{
		\cmdc \tra{\acb} \cmdc'
		\quad
		\aca \ro \fwd{\aca}{\acb}
	}{
		\prefixac \ttra{\fwd{\aca}{\acb}} \prefixacp
	}
	~~~(b)
}
\ruledefNamed{30mm}{Choice}
{choice}{
	\begin{array}{lcl}
	\cmdc \choice \cmdd &\tra{\tau}& \cmdc
	\\
	\cmdc \choice \cmdd &\tra\tau{}& \cmdd
	\end{array}
}

\vskip 2.5mm

\ruledefNamed{100mm}{While}
{while}{
	\WHbc \tra{\tau} ~ \If b \Then (\cmdc \scomp \,\, \WHbc) \Else \Skip
}

\ruledefNamed{62mm}{Locals}
{locals-reg}{
	\Rule{
		\cmdc \ttra{r \asgnlbl v} \cmdc'
	}{
		\locals{\sigma}{\cmdc}
		\tra{\tau}
		\locals{\Update{\sigma}{r}{v}}{\cmdc'}
	}
}
\ruledefNamed{56mm}{Locals/store}
{locals-store}{
	\Rule{
		\cmdc \ttra{x \asgnlbl r} \cmdc'
		\quad
		\sigma(r) = v
	}{
		\locals{\sigma}{\cmdc}
		\ttra{x \asgnlbl v}
		\locals{\sigma}{\cmdc'}
	}
}

\ruledefNamed{62mm}{Locals/load}
{locals-load}{
	\Rule{
		\cmdc \ttra{r \asgnlbl x} \cmdc'
	}{
		\locals{\sigma}{\cmdc}
		\ttra{\Readxv}
		\locals{\Update{\sigma}{r}{v}}{\cmdc'}
	}
}
\ruledefNamed{56mm}{Locals/guard}
{locals-guard}{
	\Rule{
		\cmdc \ttra{\guardlble} \cmdc'
	}{
		\locals{\sigma}{\cmdc}
		\ttra{\guardlbl{e_{\sigma}}}
		\locals{\sigma}{\cmdc'}
	}
}

\ruledefNamed{75mm}{Parallel}
{pl}{
	\Rule{
		\prp_1 \tra{\aca} \prp_1'
	}{
		\prp_1 \pl \prp_2 \tra{\aca} \prp_1' \pl \prp_2
	}
	\quad
	\Rule{
		\prp_2 \tra{\aca} \prp_2'
	}{
		\prp_1 \pl \prp_2 \tra{\aca} \prp_1 \pl \prp_2'
	}
}

\ruledefNamed{65mm}{Globals/store}
{globals-store}{
	\Rule{
		p \ttra{x \asgnlbl e} p'
	}{
		\globals{\sigma}{p}
		\tra{\tau}
		\globals{\Update{\sigma}{x}{\evalse}}{p'}
	}
}
\ruledefNamed{52mm}{Globals/load}
{globals-load}{
	\Rule{
		p \ttra{\guard{x = v}} p'
		\quad
		\sigma(x) = v
	}{
		\globals{\sigma}{p}
		\tra{\tau}
		\globals{\sigma}{p'}
	}
}

\caption{Semantics of the language}
\label{fig:semantics-main}
\end{figure}

A relevant subset of the operational rules are given in \reffig{semantics-main}.
Transitions are labelled with the syntax of the transition, i.e., assignments and guards, with the addition of the \emph{silent} label $\tau$, modelling
an internal step of a process with no effect on the context.
For brevity and ease of explanation we tend to focus on rules involving guards of a particular form, $\guard{x = v}$, which represents a load of $x$ when $x = v$.
The more general rules are given in \cite{FM18}.
We omit some rules, such as terminating rules like $\locals{\sigma}{\Skip} \tra{\tau} \Skip$. 

\refrule{reorder-rule-2} is the key rule that allows later instructions to happen earlier, according to an architecture-specific reordering relation $\ro$.  For instance,
for TSO, the main part of the reordering relation is that loads can come before stores, \ie, $x \asgn 1 \ro r \asgn y$, while $\aca \nro \acb$ for all other instruction
types. Relations for TSO, ARM and POWER are given in \cite{FM18}.  To avoid distraction in this paper we assume sequential consistency, i.e., $\aca \nro \acb$ for the
basic instruction types,
with the exception that $\tau$ steps can be reordered (allowing future local calculations to be executed ahead of time).
In \refrule{reorder-rule-2} the notation $\fwd{\aca}{\acb}$ accounts for \emph{forwarding}, where in a case such as $x \asgn 1 \cbef r \asgn x$ the instruction $r \asgn x$ can take effect
before $x \asgn 1$ provided the value 1 is forwarded to $r$, meaning that $r \asgn 1$ is executed (rather than $r \asgn x$, which it would not be sensible to execute before
$x \asgn 1$ from a sequential semantics perspective).  Forwarding is defined straightforwardly in \cite{FM18}, and we do not repeat it here.
The semantics for true prefixing, $\aca \bef \cmdc$, is given by an equivalent version of \refrule{reorder-rule-2}(a).

\refrule{choice} is straightforward for nondeterministic choice.  In \refrule{while} we unfold a loop into a conditional; the definition of conditional in a speculative
context is crucial and is deferred until \refsect{semantics-cache}.  \refrule{locals-reg} covers the case of some change to the local registers.  This
is an internal step of the process and is a silent $\tau$ step at the global level.  
\refrule{locals-store} applies when a store $x \asgn r$ is executed by a process: the local value $v$ for $r$ is substituted so that the label $x \asgn v$ is 
promoted to the global state (this rule can be generalised to cover any assignment of the form $x \asgn e$ \cite{cdsos-jlap}).
\refrule{locals-load} states that when a load $r \asgn x$ instruction is executed internally it becomes a load of $x$, i.e., a guard $\guard{x = v}$, for any value $v$.
Although there is a transition $\guard{x = v}$ for every possible $v$, only the guard with the correct value for $x$ will be possible at the system level
(via \refrule{globals-load}).
The loaded value becomes the new value for $r$ in the local state.
\refrule{locals-guard} states that a guard is evaluated with respect to the registers, and is promoted for evaluation with respect to the global state.
\refrule{pl} gives the usual interleaving model of concurrency.
\refrules{globals-store}{globals-load} straightforwardly update and access the global store via promoted stores (\refrule{locals-store}) and loads 
(\refrule{locals-load}).%
\footnote{
In this paper we assume a multicopy atomic storage system; for memory models which lack this (e.g., POWER) the storage system described in \cite{FM18} may be used.
}



Refinement ($\refsto$) is defined so that $\cmdc \refsto \cmdd$ iff all terminating traces of $\cmdd$ are also traces of $\cmdc$, 
ignoring subsequences of internal ($\tau$) steps.
Terminating traces are
those retrieved from the operational semantics where eventually $\Skip$ is reached.
(For simplicity we ignore non-terminating behaviours, that is, for this paper we consider only partial correctness, which is sufficient for detecting Spectre-like attacks.)
Note that if a behaviour is blocked (no rules are applicable, e.g., a false guard)
it 
is not considered terminating.  This eliminates behaviours where the wrong branch is incorrectly taken (as opposed to incorrectly speculated), as discussed in more detail in \cite{FM18}.

\newcommand{\horizontalEqns}[2]{
\noindent\begin{minipage}[b]{.42\linewidth}
\begin{eqnarray}
  #1
\end{eqnarray}
\end{minipage}%
\noindent\begin{minipage}[b]{.56\linewidth}
\begin{eqnarray}
  #2
\end{eqnarray}
\end{minipage}
}

We lift reasoning from the operational to refinement level via \reflaw{sos->refsto},
which allows us to straightforwardly derive \reflaw{reorder1}.
More specific laws may also be straightforwardly derived, such as resolving nondeterminism via \reflaw{choice1},
and 
\reflaw{locals-example} that hides local effects, exposing a process's global effect; this helps later to abstract from the details of transient
speculative contexts.  
\\
\horizontalEqns{
	\cmdc \tra{\aca} \cmdc'
	&\entails&
	\cmdc \refsto \aca \bef \cmdc'
\label{law:sos->refsto}
\\
	c_1 \choice c_2 &\refsto& c_1
	\label{law:choice1}
}{
	\aca \cbef \cmdc
	&\refsto&
	\aca \bef \cmdc
\label{law:reorder1}
\\
	\locals{\{r \mapsto 1\}}{x \asgn r} &\refsto& x \asgn 1
	\label{law:locals-example}
}

\section{Caches in weak memory models: \crowsl}
\label{sect:crowsl}

From the perspective of functional correctness, speculative execution may be ignored:
in the case where a 
process speculates along the branch that is eventually taken (after the conditional is evaluated) implementations ensure
that speculated instructions are committed in a consistent order;
and when speculation was down the incorrect branch any speculative computation is discarded.
However, as revealed by Spectre and other vulnerabilities, incorrect speculation can have side effects, and in this section we extend \wmml to 
expose them.

For convenience we call the extended language \crowsl, which defines conditionals to expose (incorrect) speculative execution, 
and records operations on the cache in a global variable.
Speculation occurs within a \emph{transient context}, which is discarded if speculation is found to be incorrect.

\vspace{-1.5mm}
\subsection{Syntax of \crowsl}
\labelsect{cache-syntax}
\newcommand{\ppe}{partial pre-execution\xspace} 
\newcommand{\Ppe}{Partial pre-execution\xspace} 

\subsubsection{Speculative execution.}

We introduce three new commands to capture speculative execution in \wmml.
\begin{eqnarray}
	\aca &\ttdef& \ldots  \csep \specfence
	\\
	\cmdc &\ttdef& \ldots  \csep \Speculatec \csep \Interrupt{\cmdc_1}{\cmdc_2} \csep \buffer{\sigma}{c}
	\\
	\transCtxc
	&\sdef&
	\buffer{\ess}{\locals{\sigma}{c}}
	\label{eqn:defn:transCtx}
	\\
	\IFbc &\sdef&
		\Interrupt{\Speculate{\transCtx{\cmdc_2}}}{(\guard{b} \cbef c_1)}
		~
		\choice
		~
		\Interrupt{\Speculate{\transCtx{\cmdc_1}}}{(\guard{\neg b} \cbef c_2)}
		\qquad
	\label{defn:IFSpec}
\end{eqnarray}
The instruction type $\specfence$ blocks load speculation; this is an abstract command type that may correspond to, for instance,
the \T{LFENCE} command of Intel architectures \cite{Intel64ArchManual}.  We include it to demonstrate the relevance of reordering relations and how mitigation techniques can be considered in our framework.
A \emph{speculation command} $\Specc$ gives the effect of executing command $c$ speculatively, that is, no effects on the global or local state can be seen, however, there can
be cache side effects based on the steps of $\cmdc$.  
A \emph{\ppe command} $\Intcc$ 
partially executes $\cmdc_1$ before $\cmdc_2$ begins.  The initial command $c_1$ may not execute at all, execute to completion, or partially execute.
It is the well-known CSP ``interrupt'' operator, but we rename it in this context to avoid confusion with hardware interrupts.
The \emph{transient buffer command} $\buffer{\sigma}{c}$ is used to keep track of modifications to globals executed speculatively.

We also introduce the abbreviation $\transCtxc$ which creates the \emph{transient context} for a speculative execution of $\cmdc$, that is, a (temporary)
mapping of (all) registers, and an initially empty transient buffer \refeqn{defn:transCtx}.
The values for the speculative copy of registers $\sigma$ created here is left unconstrained and may differ to the actual local state in the outer context; this accounts for
different strategies that different architectures may take.  Because the specifics of the local state are not relevant for reasoning about Spectre we do not model a
specific strategy, which could be given by adding an explicit transition that sets up the local state according to the current context.
A speculative execution of code $c$ is of the form $\Speculatenp{\buffer{\sigma_b}{\locals{\sigma_l}{c}}}$, where a copy of the locals is encapsulated in $\sigma_l$,
stores to globals are encapsulated in $\sigma_b$, and the outer speculative command generates the cache side effects.  An example of how they interact is given in
\refsect{example}.

Speculation is evident at branch points,
and hence we model conditionals differently.  Whereas in \cite{FM18}
a conditional $\IFbc$ was defined in the standard way as
$(\guard{b} \cbef \cmdc_1) \choice (\guard{\neg b} \cbef c_2)$ here we extend the definition to potentially pre-execute speculation down the
alternative branches as given in
(\ref{defn:IFSpec}).
This says there are two possibilities: speculatively execute the second branch (ignoring the guard) up until the point where the first branch is chosen, or speculatively
execute the first branch until the point the second branch is chosen.  
\OMIT{This approach of taking a speculative copy of the state, modifying the copy, and throwing it away if incorrect or overwriting the ``true'' registers if correct,
contrasts with an alternative concept which is to take a snapshot of the current state when speculation starts, then using the ``true'' registers for speculative
calculation; if speculation was down an incorrect path the snapshot it restored, and otherwise computation continues.  We may also encode this alternative approach in \wmml.
}
These two possibilities cover all behaviours relevant in the context of Spectre;
as far as is known speculation down the eventually correct branch has no impact on the security of the system
that is not already visible through other analysis techniques, e.g., information flow \cite{mur18}.
However, speculation down the correct branch is straightforward to capture, as discussed in
\refappendix{correct-branch-speculation}.

To explain the relevance of the transient context (initialised in \refeqn{defn:transCtx}) consider the execution of 
$
	\Speculate{
	x \asgn 1 \cbef r \asgn x \cbef \ldots
	}
$.
The effect of $x \asgn 1$ must not be seen globally (as it is difficult to unwind), however during speculation $r$ must use the value 1.  If instead $r$ was to
use a value of $x$ loaded from main memory this would violate local consistency (see \cite{HerdingCats}).  This detail is especially important if $r$ is used in later (speculated)
calculations, including future branches.
In our approach it emerges from the semantics that $x$ is not loaded nor drawn into the cache during speculation of the above code.
A purely syntactic approach to determining the effect of speculative execution might conclude that $x$ is added to the cache,
and hence could be overly pessimistic from a security analysis
perspective.  

\OMIT{
To explain the relevance of the transient context (initialised in \refeqn{defn:transCtx}) consider the speculative execution of the following code, which contains the potential for
a further (nested) speculation.
\[
	x \asgn 1 \cbef r \asgn x \cbef \If r > 0 \Then r_2 \asgn y \Else r_2 \asgn z
\]
The effect of $x \asgn 1$ must not be seen globally (as it is difficult to unwind), however the value of $x$ controls the later choice of branch in the conditional.
A purely syntactic approach to determining the effect of speculative execution might conclude that both $y$ and $z$ are added to the cache, which
is overly pessimistic from a security analysis
perspective.  Hence a semantic analysis is required to, for instance, avoid too many false positives.
}

Nested speculation, which may arise from nested conditionals or a speculated loop, is straightforward in our framework; a new, nested, transient context is created, and 
if an inner speculation attempts to load a global which the outer speculation has buffered
then the cache effect is removed (see \refrule{buffer}(e)).  


\subsubsection{The cache.}

The cache is modelled as a single global variable \cache, kept in the shared state, which holds a set of type $Addr$, representing addresses
(for this work we do not care what values are in the cache; however it
is straightforward to modify the type of \cache).  
We assume an uninterpreted function $\getAddr: Var \fun Addr$ such that $\getAddr{x}$ returns the address of the (global) variable $x$.
We introduce three operations on $\cache$ to model cache side channels abstractly:
cache fetching (adding something to the cache),
cache clearing (clearing the (entire) cache), 
and cache querying (checking if an address is in the cache).
Other explicit cache
operations could be added, but these are sufficient for 
modelling the attack patterns utilised to instrument Spectre attacks \cite{spectre2019}.
\begin{eqnarray}
	\cacheplusx &\sdef& \cache \asgn \cache \union \{\getAddr{x}\} 
	\label{eqn:cacheplusx}
	\\
	\clflush &\sdef& \cache \asgn \ess
	\\
	x \in \cache &\iff& \getAddrx \in \cache 
\end{eqnarray}
As these are abbreviations for updates to and guards on a global variable they fit in with the framework introduced in \refsect{background}.  
A cache fetch represents the side effect of a speculated load.
The instruction \clflush captures abstractly flushing as well as eviction of particular
cache lines as it ensures that a certain address is not present in the cache any more.

\newcommand{\cachecontext}[2]{(\textbf{cch}~#1 @ #2)}

The variable $\cache$ is kept in the global state
and hence is shared between all processes.  An alternative
would be to explicitly model it as a separate construct, \eg, $\cachecontext{C}{c}$ where $C$ is a set of addresses.  
This approach would allow more fine-grained control over cache levels, e.g., each process
could have its own L1 cache, with some subset sharing an L2 cache, with the L3 cache at the top level.
\[
	\cachecontext{L3}{\cachecontext{L2_a}{\cachecontext{L1_1}{p_1} \pl {\cachecontext{L1_2}{p_2}}} \pl 
		\cachecontext{L2_b}{\ldots}}
\]
We are interested in the worst case behaviour of the cache, where it leaks private information, 
and are not concerned with the specifics of how that may happen.  
However
details of the cache, such as its update policy,
may also be captured with extra machinery.
In that sense our model of the cache is an abstraction of the underlying 
microarchitecture implementation, which could be verified using
data and action refinement techniques \cite{HeHoareSanders-DR,Morgan-Gardiner:90,TraceReftActionSystems}.

\subsection{Semantics of \crowsl}
\labelsect{cache-semantics}
\label{sect:semantics-cache}


\subsubsection{\Ppe.}

The semantics of a \ppe process is based on that of the interrupt operator from CSP \cite{CSP}.

\vspace{1.5mm}
\ruledefNamed{110mm}{\Ppe}
{interrupt}{
	\Rule{
		c_1 \tra{\aca} c_1'
	}{
		\Intcc \tra{\aca} \Interrupt{c_1'}{c_2}
	}
	~~(a)
	\qquad \qquad
	\Rule{
		c_2 \tra{\aca} c_2'
	}{
		\Intcc \tra{\aca} c_2'
	}
	~~(b)
}\\
For commands of the form
$\Interrupt{\Specc}{d}$ the speculation of $c$ occurs for some period of time (\refrule{interrupt}(a)) before discarding the computation
and starting down the $d$ branch (\refrule{interrupt}(b)).
The arbitrariness of when $c_2$ starts captures the unknown time at which speculation may be found to be incorrect.
We make use of
the following law that covers the interruption occurring after a single action.
\begin{eqnarray}
	\Interrupt{(\aca \bef \cmdc_1)}{\cmdc_2} \refsto \aca \bef \cmdc_2
	\labellaw{intac}
\end{eqnarray}

\OMIT{
Interrupts in a sequential composition are straightforward to convert to the normal form of \wmml, \ie, 
\[
	(\Interrupt{c_1}{c_2}) \scomp c_3 = 
	\Interrupt{c_1}{(c_2 \scomp c_3)}
\]
\robnote{Actually this is more subtle; the code \emph{after} an $\If$ stmt can also be speculated.  So this should be more like
$
	(\Interrupt{c_1}{c_2}) \scomp c_3 = 
	\Interrupt{(c_1 \scomp c_3)}{(c_2 \scomp c_3)}
	$.
This actually is how it looks if you flatten ifs before doing the expansion into $\Specc$; so actually maybe this isn't needed.
}
}

\subsubsection{Transient buffers.}

Transient buffers catch stores and record them in a state; recorded values may be used for speculative computations.

\vspace{1.5mm}
\ruledefNamed{0.95\textwidth}{Buffer}
{buffer}{
	\qquad\qquad
		\qquad\qquad
	\Rule{
		c \ttra{x \asgnsmall v} c'
	}{
		\buffersc
		\tra{\tau}
		\buffer{\Update{\sigma}{x}{v}}{c'}
	}
	~~(a)
	\\ ~\\
	\Rule{
		c \ttra{\guard{x = v}} c'
		\quad
		(x \mapsto v) \in \sigma
	}{
		\buffersc
		\tra{\tau}
		\buffers{c'}
	}
	~(b)
	\quad
	\Rule{
		c \ttra{\guard{x = v}} c'
		\quad
		x \notin \dom(\sigma)
	}{
		\buffersc
		\ttra{\guard{x = v}}
		\buffers{c'}
	}
	~(c)
	\\ ~\\
	\Rule{
		c \ttra{\cacheplusxsmall} c'
		\quad
		x \in \dom(\sigma)
	}{
		\buffersc
		\tra{\tau}
		\buffers{c'}
	}
	~(d)
	\quad
	\Rule{
		c \ttra{\cacheplusxsmall} c'
		\quad
		x \notin \dom(\sigma)
	}{
		\buffersc
		\ttra{\cacheplusxsmall}
		\buffers{c'}
	}
	~(e)
}\\
\refrulea{buffer} states that (speculated) stores are recorded in the transient buffer;
\refruleb{buffer} states that (speculated) loads are serviced by the buffer (similar to \emph{forwarding} \cite{FM18}) if a value is available;
\refrule{buffer}(c) states that otherwise the load is promoted
(to be handled by the global state via \refrule{globals-load}).
In cases where nested speculation has resulted in a cache fetch,
\refrule{buffer}(d), similarly to \refrule{buffer}(b), hides a fetch of $x$ if a store of $x$ is in the buffer already;
\refrule{buffer}(e) states that otherwise the cache fetch is promoted.
In addition a transient buffer command promotes other instruction types not covered above (e.g., $\tau$, $\specfence$), and the rules do not need to cover
registers since the transient buffer encloses a local state \refeqn{defn:transCtx}.

\subsubsection{Speculation (down an incorrect path).}

Speculation should have no observable effect on registers or globals (the ``CPU state''), however in reality it may leave a footprint in the cache.
The main concept is to make explicit a cache fetch with each speculated load.

\vspace{1.5mm}
\ruledefNamed{0.95\textwidth}{Speculative context}
{speculation}{
	\Rule{
		c \ttra{\guard{x = v}} c'
	}{
		\Specc \ttra{\cacheplusxsmall}
		(\guard{x = v} \cbef \Speccp)
	}
	~(a)
	\qquad
	\Rule{
		c \tra{\tau} c'
	}{
		\Specc 
		\tra{\tau}
		\Speccp
	}
	~(b)
	\\
	\Rule{
		c \ttra{\cacheplusxsmall} c'
	}{
		\Specc \ttra{\cacheplusxsmall}
		\Speccp
	}
	~(c)
	\qquad
	\qquad
	\Speculate{\Skip} \tra{\tau} \Skip
	~~(d)
}\\
\refrulea{speculation} states that speculated loads of global variables have an initial side effect on the cache.
The load is delayed until after the cache fetch.
\refruleb{speculation} states that speculative execution can perform local computation.
\refrulec{speculation} states that cache fetches are promoted (from nested speculation).
\refruled{speculation} states that speculation may silently complete.
By omission, i.e., since there is no corresponding rule, speculation is blocked if $c$ executes a $\specfence$ command.
We do not need to consider further action types,
since speculation always encompasses a transient context out of which only loads and cache fetches are exposed.

\vspace{-2.5mm}
\subsubsection{Reordering of cache instructions.}

\newcommand{\branchF}{{\sf branch_F}}
\newcommand{\branchT}{{\sf branch_T}}
\newcommand{\rest}{{\sf branch'_T}}
\newcommand{\restp}{{\sf branch''_T}}

\newcommand{\cacheabbrev}{\textsc{c}}
\newcommand{\ca}{\cacheabbrev}

The semantics of \wmml is instantiated for a particular memory model by defining the relation $\ro$, as used in \refrule{reorder-rule-2}(b).
We must therefore define the cases under which the new (cache-based) instruction types can be reordered.  The concept of speculative execution is that loads can be 
initiated ahead of
time, though they must still (appear to) conform to the particular memory model.  However the cache fetches are not so constrained. We therefore allow cache fetch instructions to be
reordered before the majority of instruction types.
\begin{eqnarray}
	&
	y \asgn e \ro \cacheplusx \quad \mbox{iff $x$, $y$ distinct} 
	&
\labeleqn{ro-l:cacheplus}
	\\
	&
	\specfence \nro r \asgn x
	\qquad
	\specfence \nro \cacheplusx
	&
\labeleqn{ro-specfence}
\end{eqnarray}
Equation \refeqn{ro-l:cacheplus} states that 
a cache fetch of $x$ may occur earlier than loads, and stores of other variables
(note that $x \asgn 1 \nro \cacheplusx$ as the assignment will service the corresponding load, rather than memory).
Equation \refeqn{ro-specfence} states that $\specfence$ instructions block loads and cache fetches.
A potential
mitigation for the Spectre vulnerability (short of turning off speculation entirely) is to insert (concrete) $\specfence$ instructions at the start of each potentially affected branch.
However, this would have too great an impact on processor speed to be seriously considered as a blanket fix \cite{mci2019}. 

As an example of out-of-order execution with cache side effects
consider a command of the following form, where $l_i$ are loads and $s_i$ are store instructions to distinct locations.
\[
	l_1 \cbef l_2 \cbef (\If b \Then l_3 \cbef s_1 \Else s_2 \cbef l_4)
\]
Speculation allows cache fetches to come earlier (out-of-order), although whether the loads themselves can come earlier than preceding loads depends
on the architecture; ARM and POWER allow loads to be reordered, whereas TSO doesn't \cite{x86-TSO,HerdingCats}.  
Let $\ca_3$ be the cache fetch corresponding to load $l_3$.  One possible behaviour, where the $true$ branch is speculated before the $false$ branch is executed, is given by the following
sequence, which
exposes the cache fetch for $l_3$.
\[
	\ca_3 \bef l_1 \bef l_2 \bef l_3 \bef \guard{\neg b} \bef s_2 \bef l_4
\]
The cache fetch for $l_3$ occurs before the earlier loads, which, for some execution- and architecture-specific reason, have taken longer to resolve.
Note $l_3$ itself occurs in an order consistent with the memory model.  

\OMIT{
As given in \refeqn{ro-l:cacheplus} we assume an earlier store to $x$ that has not been executed blocks speculative loads of $x$, otherwise an incorrect, i.e., earlier, value for 
$x$ may be retrieved.
As a result, in the command $x \asgn 1 \scomp \If b \Then r \asgn x \scomp \ldots \Else \ldots$, the address of $x$ is not added to the cache even if
the first branch is speculated.
}

For simplicity we enforce ordering on cache operations, though the framework is flexible (for instance, on Intel architectures cache flush instructions do not necessarily prevent
\emph{pre-fetching} \cite{Intel64ArchManual}).
\[
	\alpha \nro \clflush 
	\qquad
	\clflush  \nro \alpha
	\qquad \qquad
	x \in \cache \nro \alpha
	\qquad
	\aca \nro x \in \cache 
\]
We do not intend for these to be definitive, but rather develop a framework that is flexible enough to cope with different models.

\OMIT{
Although we allow cache effects to be promoted in general, forwarding can affect them, for instance, if $x \asgn 1$ occurs (textually) immediately before a speculation
branch which loads $x$, we expect loads of $x$ to \emph{not} bring $x$ into the cache but to use that local value (as if it was in a transient buffer).  
We define a special case of forwarding from \cite{FM18} such that 
\[
	\fwd{x \asgn v}{(\cacheplusx)} = \tau
\]
That is, if the value of a global variable can be found in the pipeline, the cache is \emph{not} affected; this is again important for avoiding false positives in
information flow analysis.
}

\subsection{Example of cache side effects due to speculation}
\label{sect:example}

In this section we show the particular behaviour of a conditional statement, where the $true$ branch is (partly) speculated before the $false$ branch begins.
We construct the $true$ branch, $\branchT$, so that it modifies some global $x$ and a register $r_1$, before loading $z$ into register $r_2$ and proceeding as $\rest$.
A (partial) behaviour of $\branchT$ is given by \refeqn{branchT-trace}.
\begin{eqnarray}
	&
	\branchT \sdef x \asgn 1 \cbef r_1 \asgn 2 \cbef r_2 \asgn z \cbef \rest
		&
		\notag
	\\
		&
	\branchT \ttra{ x \asgnsmall 1 \tracesep r_1 \asgnsmall 2 \tracesep r_2 \asgnsmall z} \rest
		&
\labeleqn{branchT-trace}
\end{eqnarray}
The trace ends with a load of $z$.  We will take the case where globally $z$ has the value 42.
Now consider speculating $\branchT$.  
\begin{derivation}
	\step{
		\Speculate{\transCtx{\branchT}}
	}

	\trans{=}{Set up new transient context \refeqn{defn:transCtx}%
		\footnote{
		Note that $\branchT$ does not depend on any of the values it buffers/loads, and hence we may choose an arbitrary local $\sigma$;
		for other cases the choice of $\sigma$ may be important.}
	}
	\step{
		\Speculatenp{\buffer{\ess}{\locals{\sigma}{\branchT}}}
	}

	\trans{\tra{\tau}^*}{From \refeqn{branchT-trace}, locally update $x$ by \refrulea{buffer} and $r_1$ by \refrule{locals-reg}}
	\step{
		{\Speculatenp{\buffer{\{x \mapsto 1\}}{\locals{\Update{\sigma}{r_1}{2}}{r_2 \asgn z \cbef \rest}}}}
	}

	\trans{\ttra{\cacheplussmall{z}}}{~~~~Fetch $z$ (\refrulea{speculation}); arbitrarily assume $z$ is $42$}
	\step{
		\guard{z = 42} \cbef
		{\Speculatenp{\buffer{\{x \mapsto 1\}}{\locals{\Update{\Update{\sigma}{r{_1}}{2}}{r_2}{42}}{\rest}}}}
	}

\end{derivation}
The cache fetch has been exposed in the trace (the corresponding load $\guard{z = 42}$ is pending).  
We abbreviate the remaining code as $\restp$, and may derive \refeqn{branchT-refsto} by the above calculation and \reflaw{sos->refsto}.
\begin{eqnarray}
	&&
	\restp
	\sdef
		\guard{z = 42} \cbef
		{\Speculatenp{\buffer{\{x \mapsto 1\}}{\locals{\Update{\Update{\sigma}{r{_1}}{2}}{r_2}{42}}{\rest}}}}
		\notag
	\\
			&&
			\qquad
	\Speculate{\transCtx{\branchT}} \refsto \cacheplusz \bef \restp
	\labeleqn{branchT-refsto}
\end{eqnarray}

Now we show how the cache fetch in the $true$ branch may be seen in behaviours where the $false$ branch is taken.
\begin{derivation}
	\step{
		\If b \Then \branchT \Else \branchF
	}

	\trans{\sdef}{\refdefn{IFSpec}}
	\step{
		\Interrupt{\Speculate{\transCtx{\branchF}}}{(\guard{b} \cbef \branchT)}
		~
		\choice
		~
		\Interrupt{\Speculate{\transCtx{\branchT}}}{(\guard{\neg b} \cbef \branchF)}
	}

	\trans{\refsto}{Arbitrarily choose $false$ branch by \reflaw{choice1}}
	\step{
		\Interrupt{\Speculate{\transCtx{\branchT}}}{(\guard{\neg b} \cbef \branchF)}
	}

	\trans{\refsto}{by \refeqn{branchT-refsto}}
	\step{
		\Interrupt{
			(\cacheplusz \bef \restp)
		} {(\guard{\neg b} \cbef \branchF)}
	}

	\trans{\refsto}{by \reflaw{intac}}
	\step{
		\cacheplusz \bef 
		(\guard{\neg b} \cbef \branchF)
	}

\end{derivation}
From the system's perspective the speculation has had no effect: the assignment to $x$ was caught in the transient buffer, and then discarded, and the computations involving
registers $r_1$ and $r_2$ became silent steps that did not affect the outer state.
However, the cache has (potentially) been modified.

\newcommand{\IFbrTF}{\If b \Then \branchT \Else \branchF}

At the system level this gives the following behaviour, assuming 
global state $\sigma_g$ satisfies $\sigma_g(z) = 42$ and assuming $\sigma_g(\cache) = C$ (the value for $x$ is irrelevant), and $\sigma_l$ is the
local state (mapping $r_1$ and $r_2$).
\begin{derivation}
	\step{
		\globals{\sigma_g}{\locals{\sigma_l}{\ \IFbrTF} \pl \ldots}
	}

	\trans{\refsto}{By the above derivation (note that neither $\sigma_g$ nor $\sigma_l$ are affected)}
	\step{
		\globals{\sigma_g}{\locals{\sigma_l}{\cacheplus{z} \bef (\guard{\neg b} \cbef \branchF)} \pl \ldots}
	}

	\trans{\refsto}{Execute instruction (\refrulea{reorder-rule-2}), \refeqn{cacheplusx}, \refrule{globals-store}}
	\step{
		\globals{\Update{\sigma_g}{\cache}{C \union \{\getAddr{z}\}}}{\locals{\sigma_l}{\guard{\neg b} \cbef \branchF} \pl \ldots}
	}
\end{derivation}
The processes in `$\ldots$' could include a malicious attacker that may be able to exploit the existence of $z$ in the cache.
We give an example of this in the next section.

The derivations above cover the situation where a single speculated load is promoted to a cache fetch.  The variant of the Spectre attack we consider in the next section
contains two speculated loads; using similar reasoning to the above
we can straightforwardly show the following.
\begin{eqnarray*}
	&&
		\Speculate{\transCtx{r_1 \asgn x \cbef r_2 \asgn y}}
	\\
	&\refsto&
		\cacheplus{x} \bef \guard{x = v_1} \cbef \Speculate{\transCtx{r_2 \asgn y}}
	\\
	&\refsto&
		\cacheplus{x} \bef \cacheplus{y} \bef \guard{x = v_1} \cbef \guard{y = v_2} \cbef \Speculate{\transCtx{\Skip}}
\end{eqnarray*}
And hence by generalising \reflaw{intac} we may deduce the following.
\begin{eqnarray}
	\Interrupt{\Speculate{\transCtx{r_1 \asgn x \cbef r_2 \asgn y}}}{\cmdc}
	&\refsto &
	\cacheplus{x} \bef \cacheplus{y} \bef \cmdc
	\\
	\hskip -4mm
	\If b \Then r_1 \asgn x \cbef r_2 \asgn y \Else c & \refsto & \cacheplus{x} \bef \cacheplus{y} \bef \guard{\neg b} \cbef c
	\labellaw{if-2loads}
\end{eqnarray}

\section{Security vulnerabilities}
\labelsect{spectre}


\subsection{Attack patterns}
Cache-based timing attacks often utilise certain attack strategies to set up the cache as a covert or side channel to expose 
secret information. 
Generally, an attacker that shares a cache with a victim can observe through the variation in access time whether a particular memory address 
resides in the cache (a cache hit) and hence has been accessed previously, or not (a cache miss). 
To reduce noise on this covert channel, the attacker first ``clears'' the cache to make sure the memory address in question does not reside
in the cache. 
This can be achieved by either flushing the cache line in question (some Intel architectures offer an instruction {\tt clflush}),
or by filling the cache with other content (by accessing physically congruent addresses in a large array \cite{gru2015}), 
so that due to the contention the memory addresses in question (if present) will be evicted. 
Both these options are captured in our model through the instruction \clflush\; (as emptying the cache and filling the cache with other content
amounts to the same desired effect).

For example consider the following code that iterates over the elements of an array $B$ to determine which of $B[i]$ is in the cache.
\begin{equation}
\labeleqn{spectre-attacker}
	Atk \sdef i \asgn 0 \cbef (\While i < 256 \Do (\If B[i] \in \cache \Then r \asgn i) \cbef i \plasgn 1 )
\end{equation}
If the attacker is trying to determine the value of some byte of data $\privatedata$, then
under the assumption $B[\privatedata] \in \cache$ and for all $i \neq \privatedata$ we have $B[i] \notin \cache$ then we have
$r = \privatedata$.  

The guard $B[i] \in \cache$ is an abstraction of a timing attack that loads $B[i]$ and checks the amount of time against an architecture-specific
threshold.  For our level of analysis we do not need to explicitly model such detail, we care only that it is possible.

\FR \cite{yar2014} and also \ER \cite{gru2015}, two examples that follow the above pattern,
can be used to target the last level cache (LLC), which is shared between cores, and hence works on any cross-core as
well as cross-VM settings \cite{lip2016}. In cases where a flush instruction is not available eviction is used to ``clear'' the cache.
The following fundamental concepts of micro-architectures are exploited in these attack patterns \cite{gru2015}:
	1)
	the LLC is shared amongst all CPUs;
	2)
	the LLC is inclusive (i.e., contains all data that is stored in
  	the L1 and L2 caches, hence modifications on the LLC influence caches
  	on all other cores);
	3)
	single cache lines are shared amongst processes on the same core;
	and
	4)
	programs can map any other program binary/library into their address space.

\subsection{The Spectre attack}


\OMIT{Spectre works partly by finding code like the following in the victim ($V$).  }

Spectre attacks typically use an attack pattern based on those described above.
Additionally to setting up the cache as a channel, the attacker (mis)trains the 
branch predictor to speculate down the desired branch. 
Depending on the processor-specific branch prediction mechanism used, the training can occur by repeatedly running the code with ``correct'' input.
When unexpectedly supplied with an ``incorrect'' input, the processor will (incorrectly) speculate the desired branch,
in which secret information is loaded from memory 
(e.g., execute a memory access at an address that is
chosen by the attacker), or in other variants the attacker may leverage its own code to access the secret
from the same process, for instance, a webpage script run from within a browser process.  
In a third phase of the attack the timing difference between a cache hit and a cache
miss is observed by the attacker, as in \refeqn{spectre-attacker}, allowing it to deduce the secret value.

An example of
victim's code that is susceptible to a Spectre attack is given below (following \cite{koc2018}). 
Assume that the attacker wishes to know the value of some data $\privatedata$, held at some address in the private space of the victim process $V$
and which can be retrieved via variable $\privatevar$, \ie, $\privatedata$ is at address $\privateaddr$.  
The attacker knows/calculates the address of
$\privatevar$ relative to the victim array $A$, which we will call $\privateaddrA$, loading the value into $r_2$ via an out-of-bounds index into $A$.
That is, $A[\privateaddrA] = \privatedata$.
This private data is then used as an index into \emph{another} array $B$.  
\[
	V \sdef r_1 \asgn \privateaddrA \cbef n \asgn \#A \cbef \If r_1 < n \Then (r_2 \asgn A[r_1] \cbef r_3 \asgn B[r_2])
\]
We apply \reflaw{if-2loads} to observe the potential effects of speculation.
\begin{eqnarray*}
	V
		&\refsto&
		r_1 \asgn \privateaddrA 
		\cbef  
		n \asgn \#A
		\cbef  
		\cacheplus{A[\privateaddrA]} 
		\bef
		\cacheplus{B[\privatedata]} 
		\bef
		\guard{r_1 \not< n}
\end{eqnarray*}
(We let $\getAddr{A[i]}$ return a unique address for the array $A$ at index $i$.)
Let $\sigma_l$ be the local state for $V$ ($A, n, r_1, r_2, r_3 \in \dom(\sigma_l)$), and $\sigma_g$ the global state ($B \in \dom(\sigma_g), \sigma_g(\cache) = C$);
then we can derive the following refinement.
\[
	\globals{\sigma_g}{\locals{\sigma_l}{V} \pl \prp}
	\refsto
	\\ \qquad \qquad
	\globals{\Update{\sigma{_g}}{\cache}{C \union \{\getAddr{\privatevar}, \getAddr{B[\privatedata]}\}}}{\locals{\sigma_l'}{\Skip} \pl \prp}
\]
The data $\privatedata$ does not appear explicitly in the shared state, but indirectly through a cache fetch.
Note that the values of the variables whose addresses are in the cache are not accessible.

To infer $\privatedata$ the attacker may perform an attack as given by $Atk$ in \refeqn{spectre-attacker}.
For simplicity here we assume $Atk$ and $V$ share $B$, for instance if $B$ is a read-only array of data shared by processes in a system; alternatively $Atk$ does not need to share
$B$, but rather know where $B$ maps to in a shared cache, and map an array $B_{Atk}$ of its own so that the addresses in the cache line up.  At this level of abstraction we do not
distinguish these alternatives.
To establish the precondition that all elements of $B$ are not in the cache
the attacker sets up the context to ensure that it executes a $\clflush$ before the victim's code is run. For instance, if the vulnerable code is in a function call
provided by the server $V$, with the initial value of $r_1$ passed as an argument,
\[
	\globals{\{\cache \mapsto \_, \ldots\}}{\clflush \cbef V(\privateaddrA) \scomp Atk}
\]
This pattern can be repeated; in fact $\privateaddrA$ need not be a specific address, as data
from $V$'s private space can be read consecutively byte-by-byte by incrementing $\privateaddrA$ on each attack.


\vspace{-3.5mm}
\subsubsection{Model checking.}

We validated the semantics by encoding the refinement laws as an extension to the simulation tool described in \cite{FM18}, which is written in the Maude rewriting engine
\cite{Maude,SOSMaude}.  
The refinement laws and auxiliary definitions (such as $\ro$ for cache fetches) were encoded straightforwardly.
We then encoded the Spectre attacker and victim processes, extending the array $A$ so that its contents went beyond its stated length to model an out-of-bounds index
into private memory; the simulation runs showed 
that $r = \privatedata$ is established in the attacker \refeqn{spectre-attacker} in the cases where speculation is not interrupted in the victim until after the two 
cache fetches.

\OMIT{
\subsection{Foreshadow} \kinote{Foreshadow is a Meltdown-like attack and hence cannot be detected via code analysis.}
\subsection{Spoiler} \kinote{Spoiler just helps to speed up the derivation of the mapping from physical to virtual addresses that the attacker needs to retrieve; so is not visible in victim's code.}
}

\section{Related work}
\labelsect{relwork}

Cache side channels have been studied in the past decade (see \cite{zha2017a,ge2018} for an overview), and a number of tools 
have been developed to support the detection of vulnerabilities (e.g., \cite{doy15,wan2017,cha2018,tou2019}).
However, these developments predate the publication of the Spectre vulnerability \cite{spectre2019,koc2018} and hence do not consider
the effects of speculative execution.
Since the effects of speculation do not affect the functional correctness of an implementation (the results of incorrect speculation are thrown away),
they could be safely ignored in earlier work on the semantics of weak memory models (e.g., \cite{FM18}).
Detailed formal models of microarchitecture describe the interaction of the cache with
processors \cite{OpSemCacheCoherent}, but are not readily integrated with language-level analysis techniques.

\OMIT{ 
The formal semantics of speculative execution is a relatively new area, that has only become of significant interest since the Spectre vulnerability was detected.
Implementations of speculative execution should have no effect on semantics, since the results of incorrect speculation are thrown away, and
correct speculation should not cause more behaviours.  This allowed us to effectively ignore speculative execution in earlier work \cite{FM18}.  However the side effect of
the cache needs to be reasoned about from a software security perspective.
Detailed formal models of microarchitecture describe the interaction of the cache with
processors \cite{OpSemCacheCoherent}, but are not \kinote{readily} integrated with language-level analysis techniques.

Specific to Spectre, 
}

A model of speculative execution  
to study vulnerabilities and support the evaluation of software mitigations is presented in \cite{mci2019}.
That work assumes a uniprocessor system and is
not integrated with a weak memory model, and is designed to give a precise description of the behaviour of the microarchitecture.  
The work of \cite{dis2019} gives a model of execution that highlights speculative behaviours by explicitly modelling executions
down false branches within a partially-ordered multiset graph-based model.
In contrast to our framework, they don't consider nested speculation, nor reorder speculated instructions.

A number of tools have been developed for detecting Spectre-vulnerable code and injecting fences to mitigate the danger
\cite{oo7,ConditionalSpeculation,AbstractInterpSpecExec}
as well as information flow approaches to ensuring security in the presence of speculative execution
\cite{SPECTECTOR,FormalApproachToSecureSpeculation}.
The operational semantics underlying these approaches is 
less abstract than
that presented in this paper, and the analysis is performed at the semantic level.  
The key difference of our work is that we encode speculative execution
at the command level, and hence 
our framework supports algebraic, or refinement-based reasoning.

The CheckMate tool \cite{tri2018} integrates a model of speculative execution into a weak memory model framework \cite{PipeCheck2014}.  
Since the work aims at the verification of microarchitectures, their model is set at that level
and does not provide high-level properties such as \reflaw{if-2loads} to support reasoning on the program level.
Their tool is used to synthesise 
Spectre-style attacks and generate assembler test programs that can be used to determine if a particular processor is susceptible.
We can potentially use these test programs to
investigate the security implications 
within our more abstract framework.
We have focused on cache effects from speculative loads, however two variants of Meltdown and Spectre discovered by the CheckMate tool \cite{tri2018,SpectrePrime} 
work from speculative stores.  On architectures where speculatively executed stores affect the cache we can adapt our semantics such that 
\refrulea{buffer} emits the appropriate cache-modifying action (rather than being a purely internal step).
\vspace*{-1ex}

\section{Conclusion}
\labelsect{conclusion}
\vspace*{-1ex}

We have captured the side effects of speculative execution down the wrong path with a relatively small
extension to an existing framework for reasoning about weak memory models (out-of-order execution).  
To calculate speculated computations (beyond loads) we introduced a transient context, which is discarded in the case of incorrect speculation.
In our semantic framework,
in contrast to Plotkin-style
semantics where states appear in the configuration of the operational rules \cite{Plotkin},
we expose the effect of a transition in its label.
This simplifies semantic issues concerning redeclaration of variables
(see \cite{ifm09,cdsos-jlap} for a further discussion);
operations on variables in the inner (transient) scope become silent $\tau$ steps that do not effect the variables
in the outer scope, despite sharing the same names.
Allowing early execution of speculated instructions was straightforward to specify
in the reordering relation of \wmml \cite{FM18}.

Our intention is to allow
abstract functional analysis techniques to be used alongside security analysis techniques, reusing existing tools.  
In particular, the information flow analysis framework in \cite{mur16,mur18} has been extended to weak memory models \cite{CovernROWSL}
based on the reordering semantics of \wmml \cite{FM18}.
We envisage a further extension of that work based on \crowsl to find information leaks resulting from speculative execution.
(information flow approaches to speculative execution are also considered in \cite{SPECTECTOR,FormalApproachToSecureSpeculation}).
We have aimed to provide just enough detail so
that cache effects can be modelled, but not so much that the ability
to derive generic algebraic laws (such as \reflaw{if-2loads}) is
lost.  


\OMIT{
Cache effects can be modelled in many other ways, including with more
details of the hardware.  Such models may include models of DRAM for
``rowhammer'' attacks \cite{islam2019spoiler}, or ports, and could be
extended with real time to more accurately analyse susceptibility to
timing side channels.  
}

\OMIT{
However the most useful application of this work is in discovering new vulnerabilities.  We have highlighted some questions regarding the interplay of speculative execution
and fences, and also a further possible side effect of speculative execution on non-multicopy atomic architectures (POWER and older versions of ARM).  However it is
possible none of these are of concern, and certainly we have not developed a theory to automate an exhaustive search of all possible attackers and victims.  It is unclear
if this sort of formal modelling is suited to this task.

Theoretically this work can apply to C code using the C11 weak memory model \cite{BoehmAdveC++Concurrency}, provided the language and reordering relation can be
instantiated for the model; this is ongoing work.
}

\OMIT{ old version:
Cache effects can be modelled in many other ways, including with more details of the hardware.  Such models may include models of DRAM for ``rowhammer'' attacks 
\cite{islam2019spoiler}, or ports,
and could be extended with real time to more accurately analyse susceptibility to timing side channels.  We have aimed to provide just enough detail so that cache effects
can be modelled, but not so much that the ability to derive generic algebraic laws (such as \refeqn{ifbc-final}) is lost.  We have shown that our cache semantics can be a
relatively straightforward extension of an existing language, \wmml from \cite{FM18}, for reasoning about weak memory models, with the intention that other techniques, such as information flow
analysis, can be performed in this context.  The construction of the simulation tool means we can quickly investigate whether putative victim code can leak sensitive data.

However the most useful application of this work is in discovering new vulnerabilities.  We have highlighted some questions regarding the interplay of speculative execution
and fences, and also a further possible side effect of speculative execution on non-multicopy atomic architectures (POWER and older versions of ARM).  However it is
possible none of these are of concern, and certainly we have not developed a theory to automate an exhaustive search of all possible attackers and victims.  It is unclear
if this sort of formal modelling is suited to this task.

Theoretically this work can apply to C code using the C11 weak memory model \cite{BoehmAdveC++Concurrency}, provided the language and reordering relation can be
instantiated for the model; this is ongoing work.
Smith, Coughlin, Murray have developed an information flow checker based on the work in \cite{mur16,mur18} and extended for the semantics of \wmml, in
which we intend to extend to detect Spectre vulnerabilities.
}
{\bf Acknowledgements.}
We thank
Samuel Chenoweth, Patrick Meiring, Mark Beaumont, Harrison Cusack and the anonymous reviewers for helping us improve the paper.

\bibliographystyle{plain}
\bibliography{biblio,colvinpubs,references}

\appendix 
\section{Speculation down the correct branch; parallel speculation}
\label{appendix:correct-branch-speculation}

As far as is currently known
correct speculation has no security implications, and therefore we do not model such
behaviours explicitly.  However if needed we can capture this in several ways.   For instance,
a cache fetch can be associated with every load, whether inside or outside a speculation, similarly to \refrulea{speculation}.
Such semantics can be given by
annotating each load that may exhibit this side effect.
\[
	(r \asgn x)_{\cache} \ttra{\cacheplusxsmall} r \asgn x
\]
Alternatively we could
add the possibility of speculation down the eventually chosen branch as a choice.  
\[
		\Interrupt{\Speculate{\cmdc_2 \choice c_1}}{(\guard{b} \cbef c_1)}
		~
		\choice
		~
		\Interrupt{\Speculate{\cmdc_1 \choice c_2}}{(\guard{\neg b} \cbef c_2)}
\]
A more precise model that commits the transient context when correct speculation is found is possible, though significantly more complicated.

The concept of speculation down either branch can be extended straightforwardly to 
parallel speculation down multiple branches, for instance,
\[
	\Interrupt{(\Speculate{\cmdc_1} \pl \Speculate{c_2})}{(\guard{b} \cbef c_1)}
\]

\end{document}